\documentclass{emulateapj}
\slugcomment{Accepted for publication in ApJ Letters }

\shorttitle{DISENTANGLING THE HERCULES STREAM}
\shortauthors{BENSBY ET AL.}

\newcommand\teff{T_{\rm eff}}
\newcommand\kms{\rm\,km\,s^{-1}}

\newcommand\ulsr{U_{\rm LSR}}
\newcommand\vlsr{V_{\rm LSR}}
\newcommand\wlsr{W_{\rm LSR}}

\begin{document}

\title{
Disentangling the Hercules stream\altaffilmark{1}
}
\author{
T. Bensby,\altaffilmark{2}
M.S. Oey,\altaffilmark{2}
S. Feltzing,\altaffilmark{3}
and
B. Gustafsson\,\altaffilmark{4}
}

\altaffiltext{1}{Based on observations collected with the MIKE spectrograph 
on the 6.5\,m Magellan/Clay telescope at the Las Campanas observatory in
Chile}
\altaffiltext{2}{Department of Astronomy, University of Michigan, 
830 Dennison Building, 500 Church Street, Ann Arbor, MI 48109-1042, USA;
{\tt tbensby@umich.edu, msoey@umich.edu}}
\altaffiltext{3}{Lund Observatory, Box 43, SE-221\,00 Lund, Sweden;
{\tt sofia@astro.lu.se}}
\altaffiltext{4}{Department of Astronomy and Space Physics, University 
of Uppsala, Box 515, SE-751\,20 Uppsala, Sweden;
{\tt bg@astro.uu.se}}

\begin{abstract}
Using high-resolution spectra of nearby F and G dwarf 
stars, we have investigated the detailed abundance and age structure of the
Hercules stream. We find that the stars in the stream 
have a wide range of stellar ages, metallicities, and element abundances. 
By comparing to existing samples of stars in the solar neighbourhood with
kinematics typical of the Galactic thin and thick disks we find
that the properties of the Hercules stream 
distinctly separate into the abundance and age trends 
of the two disks. Hence, we find it unlikely that the 
Hercules stream is a unique Galactic stellar population, but rather a 
mixture of thin and thick disk stars. This points 
toward a dynamical origin for the Hercules stream, probably caused by the 
Galactic bar. 
\end{abstract}


\section{Introduction}

The stellar velocity distribution in the solar neighbourhood
is manifoldly structured 
\citep[see, e.g.,][]{dehnen1998a,skuljan1999,famaey2005,helmi2006,
arifyanto2006}. Prominent features are the Pleiades-Hyades super-cluster, 
the Sirius cluster, and the Hercules stream (also known as the $u$-anomaly). 
From a large sample of nearby G and K giants \cite{famaey2005} 
found that the Hercules stream makes up approximately 6\,\% of the stars in the 
solar neighbourhood, and that they have a net drift of 
$\sim 40\,\kms$ directed radially away from the Galactic centre. Just 
as for the Galactic thick disk, their orbital velocities around the Galaxy 
lag behind the local standard of rest (LSR) by $\sim 50\,\kms$ 
(see also \citealt{ecuvillon2006astroph} who found
similar properties for the Hercules stream using nearby F and G dwarf stars). 

Numerical simulations have shown that the excess of stars at 
$(\ulsr,\vlsr)\approx(-40,\,-50)\,\kms$ can be
explained as a signature of the Galactic bar
\citep[e.g.,][]{raboud1998,dehnen1999,dehnen2000,fux2001}. 
If it is a chaotic process, where stars gets gravitaionally scattered from the 
inner Galactic regions by the bar, or if they are coupled to the outer 
Lindblad resonance of the bar is, however, uncertain \citep{fux2001}. 
Either way, this points to an origin for the Hercules stream 
that is related to the inner disk regions. So, is the Hercules stream a
distinct stellar population with a unique origin and evolutionary history, 
or is it a mixture of different populations? 

Using available data in the literature \cite{soubiran2005} found that most 
Hercules stars tend to follow the thin disk abundance trends. They, however,
concluded that the existing knowledge about the chemical properties of the 
Hercules stream did not admit safe conclusions about the origin of the
stream. We note that the distinct 
and well separated chemical signatures that the two disks exhibit 
\citep{fuhrmann2004,bensby2003,bensby2004,bensby2005,mishenina2004,feltzing2006}
are less separated in the abundance plots in
\cite{soubiran2005}  (compare, e.g., the abundance trends for oxygen in 
their Fig.~10 with Fig.~10 in \citealt{bensby2004}). This is likely an effect
of merging different data sets, each containg different systematic
errors.

To further study the origin of the Hercules stream we have observed a sample
of 60 F and G dwarf stars, all kinematically selected to be members of the
stream. By performing a strictly differential detailed abundance analysis 
of the Hercules stream stars relative to stars of the two disk populations 
previously studied by us \citep{bensby2003,bensby2005}
we minimise uncertainties due to systematic errors in the analysis.

In this letter we focus on two elements that
show distinct abundance trends for the thin and thick disks:
Mg \citep[e.g.,][]{fuhrmann2004,feltzing2003,bensby2003,bensby2005} and
Ba \citep[e.g.,][]{mashonkina2003,bensby2005}. 
Other $\alpha$-elements, iron peak elements, and $r$- and $s$-process 
elements will be presented in an upcoming paper 
(Bensby et al.~2007, in prep.), wherein we also will 
describe the observations, data reductions, abundance analysis, etc. 
\citep[see, however,][where the results for a few other elements are briefly 
presented]{bensby2006mru}.

\section{Identification of Hercules stream stars}  \label{sec:sample}

\begin{figure}
\centering
\resizebox{\hsize}{!}{
  \includegraphics[bb=50 28 540 760, clip, angle=-90]{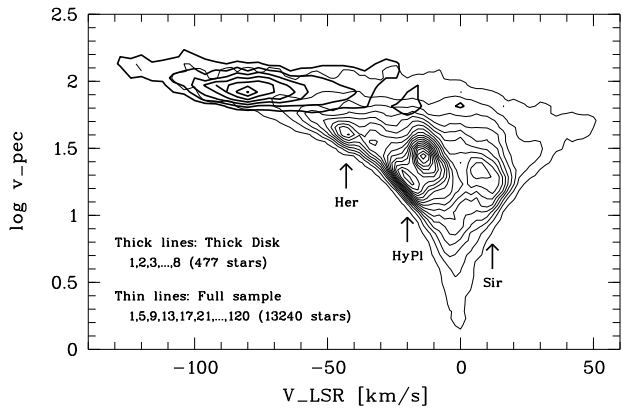}}
\caption{
        Velocity distribution of the F and G dwarf stars in the
        \cite{nordstrom2004} catalogue 
        (where $v_{\rm pec}\equiv(\ulsr^2 + \vlsr^2 + \wlsr^2)^{1/2}$).
        The Hercules stream (Her), the Hyades-Pleiades cluster (HyPl),
        and the Sirius group (Si) have been marked. Distances between the
        isodensity curves are also given in the plot.
        }
  \label{fig:contour}
\end{figure}

We used the kinematical method from \cite{bensby2003, bensby2005}
to define a sample of Hercules stream stars. This method assumes
that a stellar population has a Gaussian velocity distribution and constitutes
a certain fraction of the stars in the solar neighbourhood.
Assuming that the solar neighbourhood is a sole mixture of the thin disk, 
the thick disk, the Hercules stream, and the halo, it is then possible to 
calculate the probabilities for individual stars (with known space velocities)
to belong to either of the populations. We selected Hercules 
stream stars as those stars that have probabilities of belonging to 
the Hercules stream that are at least as large as twice the probabilities of 
belonging to any of the other populations. 

The \cite{nordstrom2004} catalogue contains kinematic 
information for 13240 F and G dwarf stars in the solar neighbourhood.
Considering the full catalogue we are able to kinematically tag 12040 stars
as likely thin disk members, 438 as likely thick disk members, and 112 as 
likely Hercules stream members. Fig.~\ref{fig:contour} shows a contour plot 
of the distribution for all stars in the catalogue (thin solid lines), 
and density contours for the thick disk (thick solid lines).

\begin{figure}
\resizebox{\hsize}{!}{\includegraphics[bb=18 165 592 543,clip]{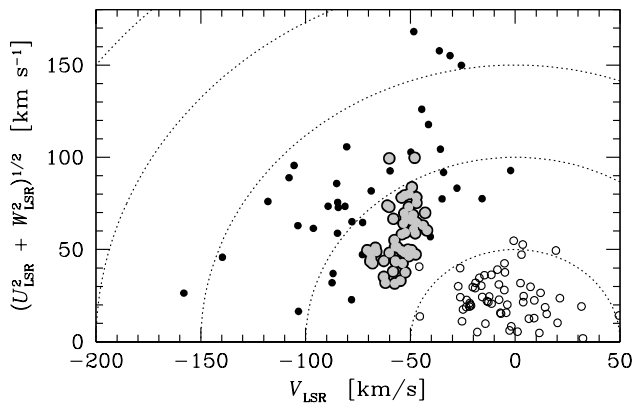}}
\caption{
        Toomre diagram for the Hercules stream sample
        (larger gray circles).  Thin and thick disk stars
        from \cite{bensby2003,bensby2005} are marked by open and black
        smaller circles, respectively.
        }
\label{fig:kinematics}
\end{figure}

The 60 Hercules stream stars in this study are 
shown in a Toomre diagram in Fig.~\ref{fig:kinematics} together
with the 102 thin and thick disk stars from \cite{bensby2003, bensby2005}. 
Our Hercules stream sample is well confined and forms a distinct kinematical
group. None of the thin and thick disk stars in \cite{bensby2003,bensby2005}
can be classified as Hercules stream stars.

\section{Brief summary of observations, abundance analysis, 
and age determinations} 
\label{sec:observations}

High-resolution ($R\approx65\,000$), high-quality ($S/N\gtrsim250$) 
echelle spectra were obtained for 60 F and G dwarfs by TB in Jan, Apr, 
and Aug in 2006 with the MIKE spectrograph \citep{bernstein2003} on the 
Magellan Clay 6.5\,m telescope at the Las Campanas Observatory in Chile. 
Solar spectra were obtained during the runs by observing the asteroid Vesta 
(in Jan), the Jovian moon Ganymede (in Apr), and the asteroid Ceres (in Aug).

For the abundance analysis we used the Uppsala MARCS stellar model 
atmospheres \citep{gustafsson1975,edvardsson1993,asplund1997}. 
The chemical compositions of the models were scaled with metallicity 
relative to the standard solar abundances as given in 
\cite{asplundgrevessesauval2005}, but with $\alpha$-element 
enhancements\footnote{The $\alpha$-element enhancement increases linearly 
from  $\rm [\alpha/Fe]=0$ at $\rm [Fe/H]=0$ up to $\rm [\alpha/Fe]=0.4$ at
$\rm [Fe/H]=-1$, and is then constant $\rm [\alpha/Fe]=0.4$ for lower 
metallicities.} for stars with $\rm [Fe/H]<0$. To determine the effective 
temperature and the microturbulence parameter we required 
all Fe\,{\sc i} lines to yield the same abundance independent of lower
excitation potential and line strength, respectively. 
To determine the surface gravities we utilised that our stars have accurate 
Hipparcos parallaxes \citep{esa1997}. Final abundances were 
normalised on a line-by-line basis with our solar values as reference and 
then averaged for each element. 

Stellar ages were determined from the Yonsei-Yale (Y$^{2}$) $\alpha$-enhanced  
isocrones \citep{kim2002,demarque2004} in the $\teff$-$M_{\rm V}$ plane.
Upper and lower limits on the ages were estimated from the error bars due 
to an uncertainty of $\pm 70$\,K in $\teff$ 
and the uncertainty in $M_{\rm V}$ due to the error in the parallax
\citep[see][]{bensby2003}.

\section{Results and discussion} \label{sec:results}

\begin{figure*}
\resizebox{\hsize}{!}{\includegraphics[bb=18 223 592 508,clip]{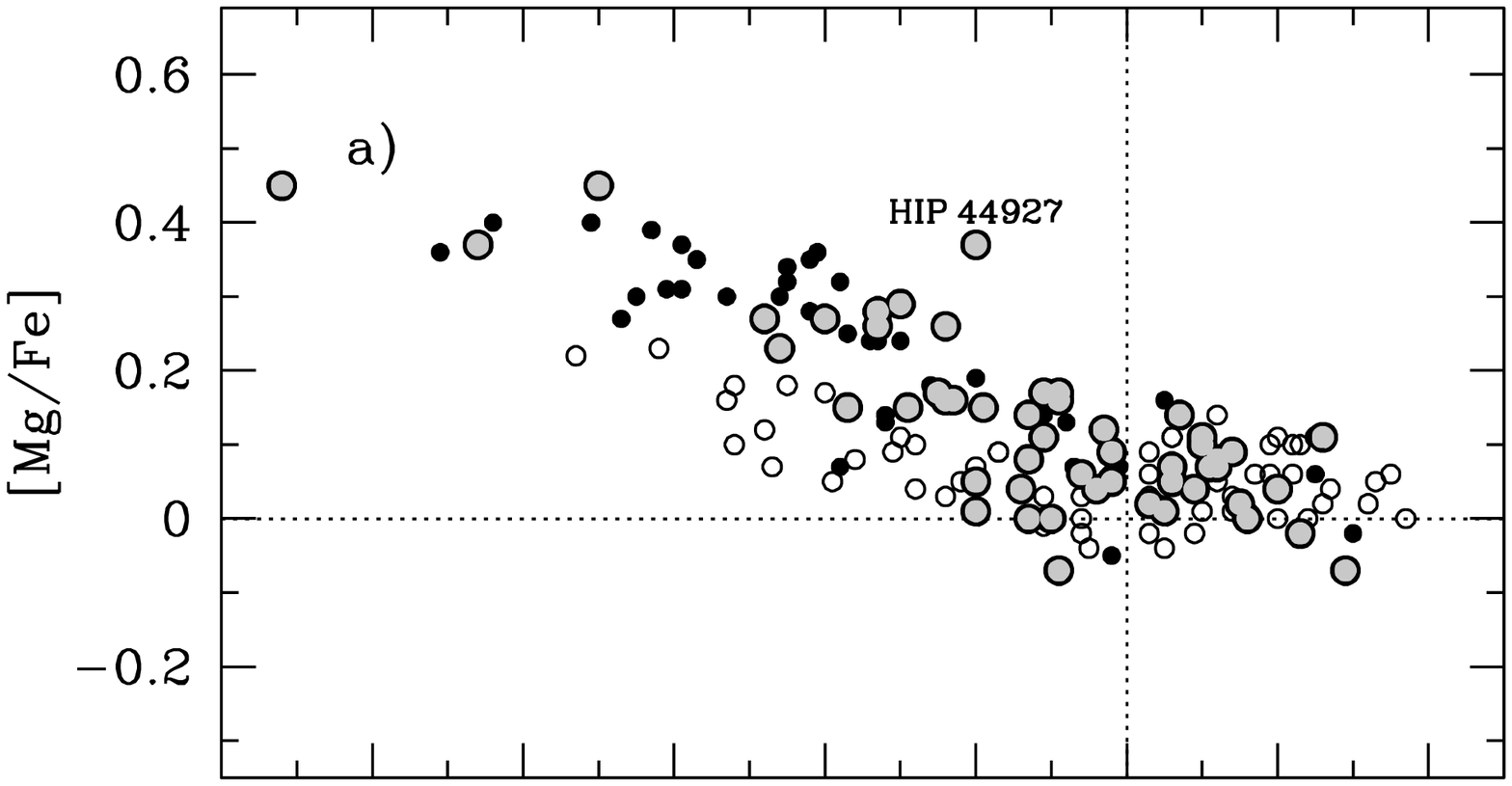}
                      \includegraphics[bb=18 223 592 508,clip]{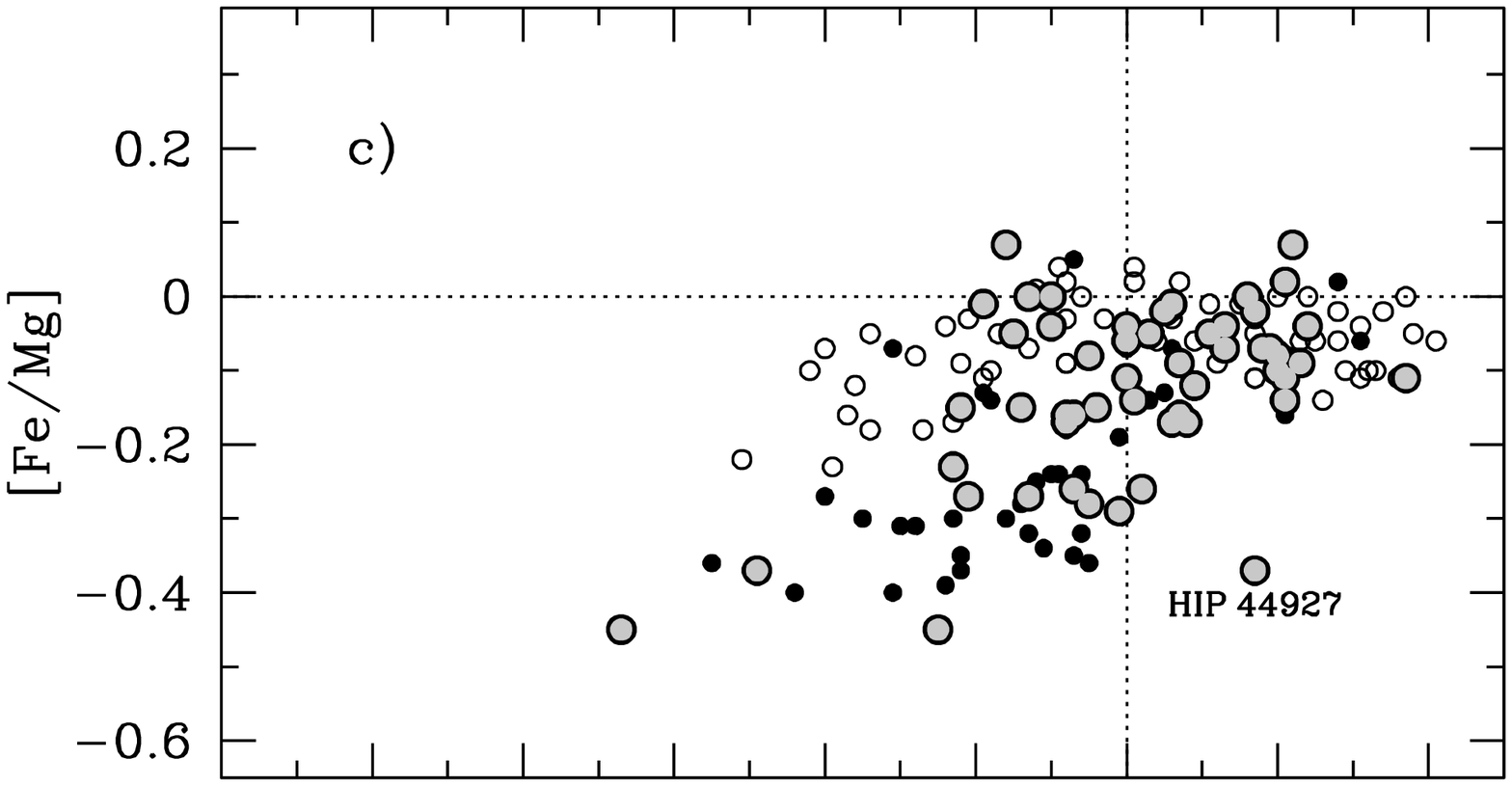}}
\resizebox{\hsize}{!}{\includegraphics[bb=18 165 592 501,clip]{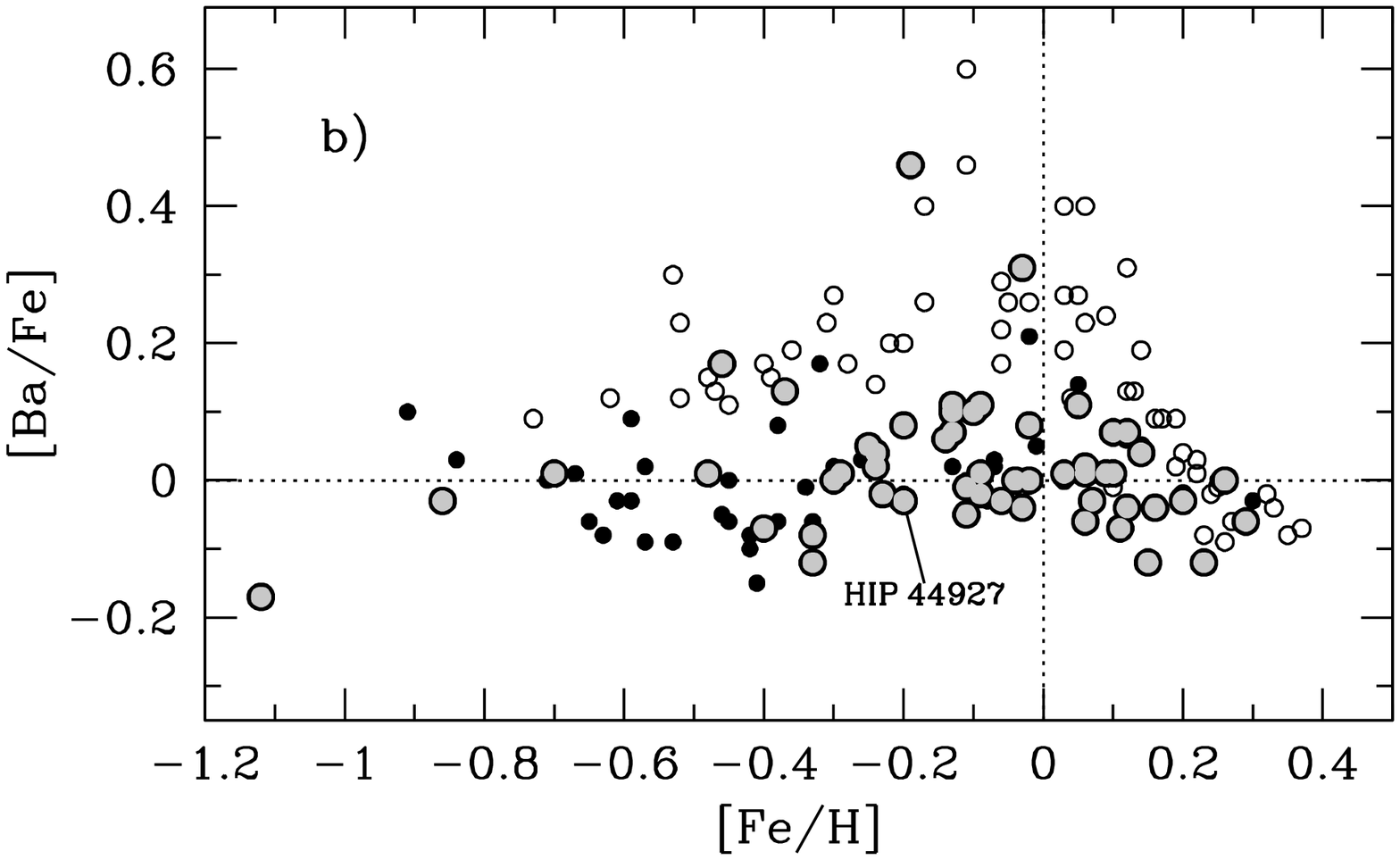}
                      \includegraphics[bb=18 165 592 501,clip]{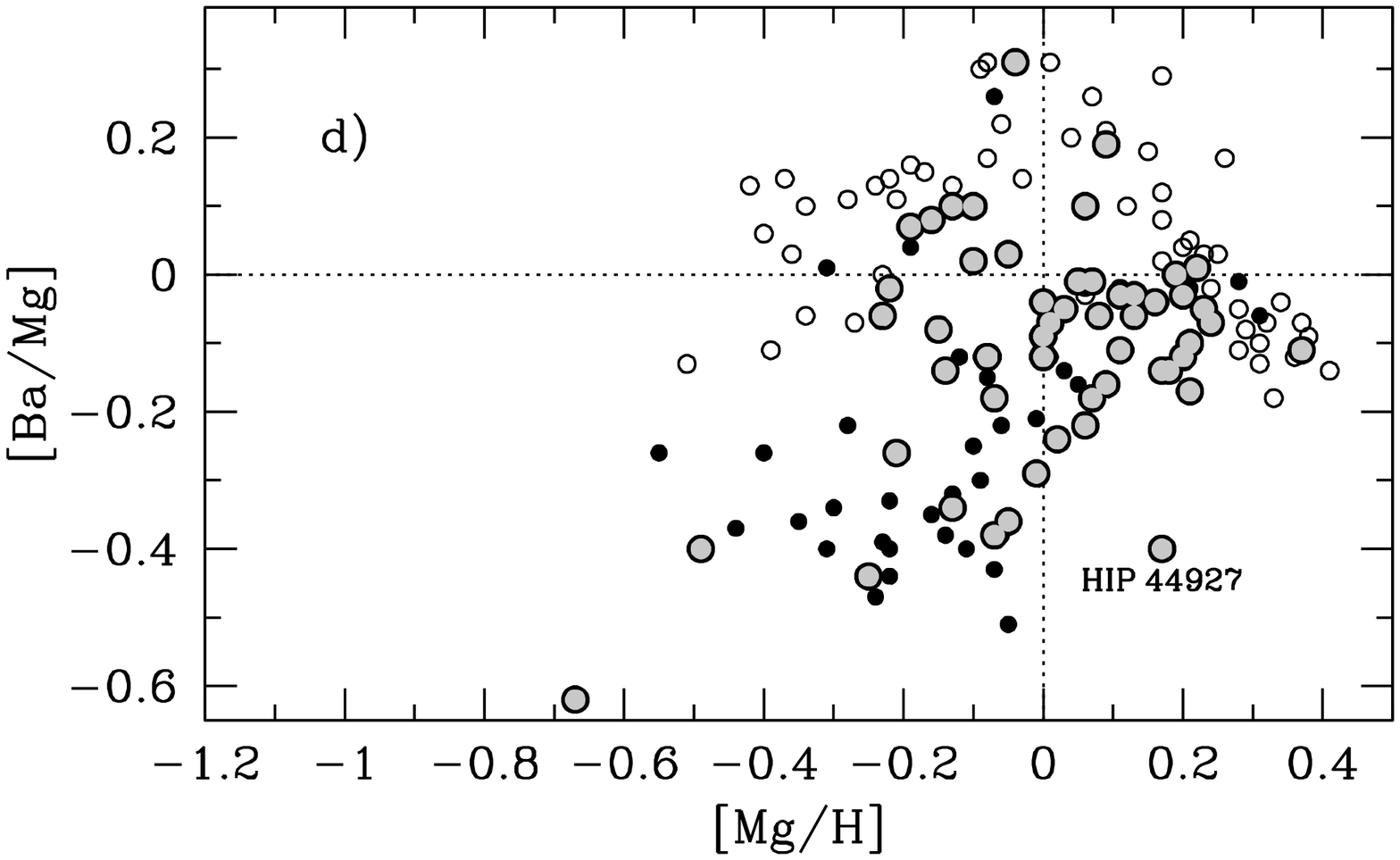}}

\caption{
        [Mg/Fe] and [Ba/Fe] vs. [Fe/H] (panels to the left) and
        [Fe/Mg]\,-\,[Mg/H] and [Ba/Mg] vs. [Mg/H] (panels to the right).
        Hercules stars are marked by larger gray circles.
        The thin and thick disk stars from
        \cite{bensby2003,bensby2005} are marked by open and black
        smaller circles, respectively.
        }
\label{fig:abundancetrends}
\end{figure*}

Our abundance results are shown in Figs.~\ref{fig:abundancetrends}a-d.
While the [Mg/Fe] vs.~[Fe/H]
trends for the thin and thick disks are clearly separated 
they do merge as $\rm [Fe/H] \approx0$ is reached.
The [Ba/Fe] vs.~[Fe/H] trend, on the other hand, keeps the two disks well
separated until $\rm [Fe/H] \approx0.1$. 
The separation between the thin and thick disks is larger when 
Mg is used as the reference element. 

The observed thick disk [Mg/Fe] vs.~[Fe/H] trend is due to 
the different production sites of Mg and Fe in SN\,II and SN\,Ia, respectively,
operating at different time scales
\citep[see, e.g.,][]{mcwilliam1997}. 

The solar system abundance of Ba has been built up by two
different production mechanisms that work on different timescales;
the $r$-process which dominated at low metallicities ($\rm [Fe/H]\lesssim-1.5$)
and contributed $\sim20$\,\%, and the 
$s$-process that dominated at higher metallicities and
contributed $\sim80$\,\% \citep{travaglio1999}.
The increase (or the ''bump") that can be seen in the [Ba/Fe]-[Fe/H] trend 
for the thin disk is likely to be a signature of when the $s$-process
enrichment from AGB stars became significant \citep{travaglio1999}. 
For other environments the mixture may be different, compare e.g.
the Ba and Eu trends in \cite{mashonkina2003}.
For the thick disk, the flat [Ba/Fe] trend
could indicate that the $s$-process did not play a significant r\^ole
in the Ba enrichment of the gas from which the thick disk formed
(and hence the lack of the ``bump").

\subsection{Abundance bimodality?}

A first impression is that most Hercules stream stars
follow the trends as outlined by the thick disk. In the 
[Ba/Fe] vs.~[Fe/H] plot this is especially evident and even up to 
metallicities as high as $\rm [Fe/H] \approx +0.1$. In the [Mg/Fe] vs.~[Fe/H] 
plot the Hercules stars follow the thick disk trend up to 
$\rm [Fe/H]\approx-0.2$ and may then show signs of a mixing between the 
two disks for higher metallicities. The [Fe/Mg] and [Ba/Mg] trends with [Mg/H]
give similar results. 

Apart from one star, HIP\,44927, we do not find any Hercules stream
stars that deviate from the thin and thick disk abundance trends.
Recent observations have shown that bulge stars have
large $\alpha$-enhancements at solar and super-solar metallicities
\citep{fulbright2006astroph,zoccali2006}. This compares very well with what
we see for HIP~44927, indicating that we might have picked up a bulge
star in our Hercules sample. Its age $3.8^{+0.7}_{-0.4}$\,Gyr is, however, 
inconsistent with the bulge being an old population. But there are
indications that the bulge could contain stars as young as a few hundred
million years \citep{gilmore2003}.

\subsection{Age bimodality?}

\begin{figure}
\resizebox{\hsize}{!}{\includegraphics[bb=18 165 592 520,clip]{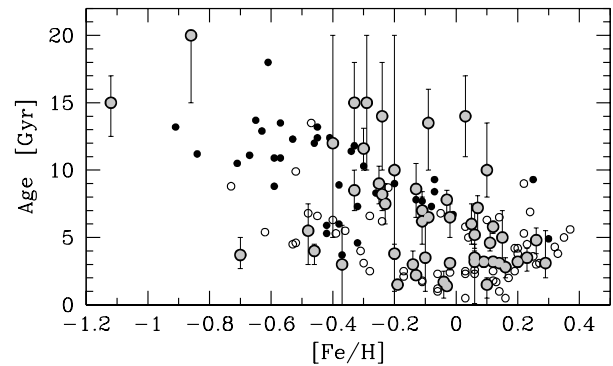}}
\caption{Ages vs. [Fe/H] for the Hercules stream stars (larger gray circles).
        Thin and thick disk stars from  \cite{bensby2003,bensby2005} are
        marked by open and black smaller circles, respectively. Error
        bars show the upper and lower age estimates due to the
        uncertainties in the $\teff$s and in the distances to the stars.
        }
\label{fig:agefeh}
\end{figure}

In Fig.~\ref{fig:agefeh} we plot stellar ages versus [Fe/H]. At metallicities
below $\rm [Fe/H]=0$ it appears that the Hercules stream divides into two
(or maybe three) branches: one that follows the thick disk age
trend (the downward age-metallicity relation we see for the thick disk
trend was seen in \citealt{bensby_amr} and then verified
by \citealt{haywood2006} and \citealt{schuster2006});
one that follows the thin disk age trend; and a
few stars (4-5) that tend to have high ages of $\sim15$\,Gyr
in the interval $\rm -0.4\lesssim[Fe/H]\lesssim0$.
Once again, we see that the stars of the Hercules stream follow
the trends outlined by the stars with kinematics typical of the
thin and thick disks.
As the uncertainties in the age determinations are notoriously
difficult to estimate \citep[see e.g.][]{jorgensen2005},
and generally come out too small when using standard methods
one should be careful to make far fetched interpretations
about the few outliers.

\subsection{The origin of the Hercules stream}
\label{sec:origin}

Some Galactic streams may have origins due to minor mergers
\citep[e.g.][]{navarro2004}. Could this be the case for the
Hercules stream? Such a merging system must then have had chemical properties
similar to the present Galactic disks. Thus, such a merging galaxy must
have had chemical properties that depart considerably from
those of local dwarf galaxies \citep[cf., e.g.][]{venn2004} and would
presumably be more similar to a major spiral galaxy. There is no 
observational evidence that such a substantial merger occured 
in the Milky Way during the last 10\,Gyr \citep{gilmore2002}.
Instead, our results strongly suggest that the Hercules stream has a dynamical
origin, presumably due to the dynamical effects of the Galactic bar,
and it may consist of stars from the inner Galactic disks.

Are we then tracing the inner thin disk or the inner thick disk?
As no detailed abundance data for the inner disks have been obtained
yet, only indirect evidence is available. The thick disk velocity 
dispersion is about twice that of the thin disk, but its rotational
velocity is just $\sim80$\,\% of that of the thin disk.
According to the Toomre stability criterion
\citep[$Q\propto V_{\rm rot} \cdot \sigma_{\rm R}$;][]{toomre1964}
the thick disk should be slightly more stable than the thin disk
($Q_{\rm thick\,disk}\approx1.6\cdot Q_{\rm thin\,disk}$) and less
disturbed by gravitational influences from the Galactic bar.
Since the inner thin disk is thought to have more evolved stellar populations,
as judged from its abundance and age gradients, the higher ages and
$\rm [\alpha/Fe]$ ratios of the Hercules stream seem consistent with an
inner thin disk origin. On the other hand, the abundance patterns are better
matched to the thick disk in general, and it is plausible that the thick 
disk also has a metallicity gradient, which would then match the relative
number of high-metallicity stars in the stream.

As the stars of the Hercules stream display the distinct abundance and age 
trends of {\it both} the thin and thick disk stars, it is likely that
it actually is a mixture of the two disks. This is also suggested 
by the Toomre stability criterion being similar for the two disk
components.
A preliminary investigation of the relationship between the ages and
abundance ratios for our stars in the Hercules stream 
shows that they are correlated in a way expected for
an Hercules stream composed of Galactic disk stars
(see further Bensby et al. 2007, in prep.)

\section{Conclusion}

The Hercules stream is
unique in the sense that it forms a well defined enhancement
in the velocity distribution of nearby stars. 
However, large ranges in stellar ages and
elemental abundances indicate that it is not a distinct
stellar population. Instead, as the age and abundance trends 
in the Hercules stream are similar to the trends in the thin and thick 
disks, we conclude that the Hercules stream is a mixture of thin and thick disk 
stars. This result is in concordance with models that suggest that the 
kinematical properties of the Hercules stream are coupled to dynamical 
interactions with the Galactic bar. Whether the Hercules stream stars 
have a real inner disk origin or whether they are nearby stars whose 
kinematics are an effect of the outer Lindlad resonance of the bar, 
which may be situated near the solar
neighbourhood \citep{dehnen2000}, is unclear. 
Further insights into this problem 
could be gained by making comparisons to detailed abundance data from the
{\it in situ} inner Galactic thin and thick disks.
As such data currently are unavailable, we have initiated a study to
obtain them.

\acknowledgments

This work was supported by the National Science 
Foundation, grant AST-0448900, and a grant from the Swedish Research Council. 
SF is a Royal Swedish Academy
of Sciences Research Fellow supported by a grant from the Knut and
Alice Wallenberg Foundation.

\end{document}